\documentclass[aps,prd]{revtex4}

\usepackage{amsmath}

\begin{document}

\title{Mirror as polaron with internal degrees of freedom}

\author{G.E.~Volovik}
\affiliation{Low Temperature Laboratory, Aalto University,
 P.O. Box
15100,
FI-00076 Aalto, Finland}
\affiliation{Landau Institute for Theoretical Physics RAS,
Kosygina 2,
119334 Moscow, Russia}

\author{M.A.~Zubkov}
\affiliation{University of Western Ontario,  London, ON, Canada N6A 5B7}
\affiliation{ITEP, B.Cheremushkinskaya 25, Moscow, 117259, Russia}

\begin{abstract}
We consider the model suggested by  Wang and Unruh \cite{WangUnruh2013}  for
the $1+1$ D mirror moving in the quantum vacuum. We consider the
relation of this model to the problem of polaron -- the electron moving in
the vacuum of the quantum field of phonons. We introduce the field -
theoretical model of such a mirror. It contains the multi - component spinor
field interacting with the scalar field. We discuss the source of the
logarithmic
divergence in the mirror mass and its relation to the problem of the
divergencies in vacuum energy.
\end{abstract}



\maketitle

\section{Introduction}

The authors of  \cite{WangUnruh2013} consider
the mirror with an internal harmonic oscillator coupled to
a scalar field in $1 + 1$ D. It is found that the
effective mass (the rest energy) of such a composite mirror is infinite
(logarithmically divergent) due to the vacuum
fluctuations of the scalar field. This system is considered in
\cite{WangUnruh2013} as the counter - example to the generally accepted
statement, that the vacuum energy matters only when taking gravity into
account (otherwise one can only measure the energy differences). It is
argued, that in the given system the infinities like that of the vacuum
energy do matter.

First of all, we disagree with the mentioned generally accepted statement.
In condensed matter physics the
vacuum energy density participates in the thermodynamics of the system
together with the energy density of matter. That is why we expect, that the
vacuum energy density
matters both in the Universe with gravity and in the Universe
without gravity (see the latter case in Ref. \cite{Volovik2003}).
When one tries to calculate the vacuum  energy density or the ground state
energy density of condensed matter systems summing the energies of
fluctuations, one obtains the divergent sum with the
utraviolet (UV) cutoff determined by the high-energy scale. In particle
physics such estimate suggests a huge value of the cosmological constant.
However, the
condensed matter systems, where the microscopic
physics (the analog of the trans-Planckian physics) is known, demonstrate
that the divergent high-energy
contributions from zero point energies of quantum field are cancelled by the
microscopic (trans-Planckian) degrees of freedom due to the thermodynamic
identities.

Let us consider (very roughly) the open system that is  described by the bosonic fields $\phi$. We suppose, that the volume $V$ of this system is variable. Also we imply the existence of the conserved charges $Q$. In quantum field theory vacuum typically corresponds to vanishing charges. However, in our case we consider the situation, in which the values of $Q$ are fixed and nonzero (for example, we may consider the system consisted of interacting particles with the fixed number $N=Q$ of the particles).
In vacuum of such a system the values of the fields $\phi$ are equal to their average values $\phi_0$. We assume, that the vacuum is homogenious, so that $\phi_0$ does not depend on the position in space. Total energy $E = \int d^3 x \epsilon[\phi_0,Q/V]$ is expressed through the energy density $\epsilon$ that depends on $\phi_0$ and on the charge density $q = Q/V$.  One of the conditions for the equilibrium is
 $\partial\epsilon[\phi_0,q]/\partial \phi_0=0$, and another one is that the variation of $E$ over the volume vanishes. This gives $\epsilon[\phi_0,q] - q\partial\epsilon[\phi_0,q]/\partial q= 0$. If the same pattern is applicable to particle physics, and to the Universe as a whole, then we are to treat the overall volume of the Universe as variable. Therefore, the equilibrium is achieved when the thermodynamic potential
 $\Omega=\epsilon[\psi,q] - q\partial\epsilon[\psi,q]/\partial q$ vanishes. It is this thermodynamic
 potential, which is equal to the cosmological constant (see \cite{KlinkhamerVolovik2010} and recent review \cite{Volovik2013}).

            Within this approach it is naturally to assume, that we live near the
equilibrium. Therefore, the total vacuum energy density  $\Omega$ with all contributions taken into account should be very close to zero. This provides the cancellation of the main UV divergent terms in the known low energy theory by something coming from the unknown high energy theory. As a result only a small fraction of
the vacuum energy density remains that is comparable with the observed
value of the cosmological constant. This cosmological constant gives rise to the accelerated
expansion of the Universe. The latter expansion should be understood as the small deviation of the Unverse from equilibrium.

A similar mechanism may work also for the cancellation of the quadratically
divergent contributions to the Higgs boson mass (see, for example,
\cite{Volovik:2012qq} and references therein). For the practical realization
of this mechanism it is necessary to point out the reason for the
cancellation (regularization) of the UV divergent contributions to the Higgs
boson mass. This may occur due to a certain physical principle and may not
be related to the details of the underlying
microscopic physics (that is in this case the physics of the UV completion
of the Standard Model).


Next, we analyze the model discussed in Ref. \cite{WangUnruh2013}. This
model operates with the massive oscillator moving along its world - line and
interacting with the scalar field. The world - line fluctuates, and its fluctuations are to be defined in
accordance with the action functional of the moving oscillator. We reformulate this
model in a standard way in terms of the quantum field of the moving
oscillator. This allows to calculate the renormalization of mass for the
given object due to the interaction with the scalar field using the standard perturbation theory.
The obtained infinity  is
the result of the unrealistic approximation made in the model: the
mirror is assumed to be infinitely thin and is approximated by the
$\delta$-function. The
finite thickness $R$ of any realistic mirror provides the physical UV cutoff
to the logarithmically divergent integral, $E_{\rm UV} \sim \hbar c/R$.

Our field - theoretical formulation is similar to the field - theoretical
description of the
Fr\"ohlich polaron.  In many respects the mirror moving in the
quantum vacuum is similar to the impurity moving in the condensed matter
"vacuum"  such as Fr\"ohlich polaron (electron moving in the vacuum of
phonon quantum field, see review \cite{Devreese2009});  Bose polaron
(impurity moving in a Bose-Einstein condensate, see Ref. \cite{LiSarma2014}
and references therein);  polaron in fermionic "vacuum" (see recent review
\cite{LanLobo2014});  gravitational polaron (see Ref.
\cite{ComechZubkov2013}), etc. However, the model discussed in Ref.
\cite{WangUnruh2013} has a peculiar property:  the interaction between the
oscillator and the scalar field is proportional to the time derivative. This
is the source of the
logarithmically divergent positive contribution to the
polaron energy, which adds to the regular negative contribution in the
conventional polaron problem  and thus reverses
the sign of the correction to the mirror mass.

\section{Non - relativistic moving oscillator}

\subsection{Fr\"ohlich - like description of moving oscillator}

Throughout the text we adopt the system of units with $c=\hbar=1$.

In this section we assume that the thickness of the mirror exceeds its
Compton length,
$R \gg 1/ M$, where $M$ is the bare mass of the mirror.
This means that $E_{\rm UV} \ll M $ and we are able to use the non -
relativistic approximation.

The classical action for the mirror interacting with the oscillator bound to
the mirror and with the massless scalar field can be written in the
following form  (see Eq.(A1) of Ref.  \cite{WangUnruh2013}):
\begin{eqnarray}
S &=& \frac{M}{2}\int\!\,dt\, \dot{y}^2(t) +\frac{1}{2}\int\!d x\,dt\,
\Big[
(\partial_t{\phi})^2
-
(\partial_x\phi)^2
\Big]\nonumber\\&& + \frac{1}{2}\int\!\,dt\,
\Bigl(\dot{q}^2(t) - \Omega^2 q^2(t) - 2
\epsilon \,\dot{q}(t) \phi(y[t],t)\Bigr)
\label{ClassicalAction}
\end{eqnarray}
Here $y(t)$ is the coordinate of a mirror; $q(t)$ is the variable of the
oscillator; and $\phi(x,t)$ is the $1+1 D$ scalar field. Here for simplicity
we assume that the bare mass $M$ of the mirror is much larger, than the
ultraviolet cutoff,  so that its motion
can be considered  in the non-relativistic approximation.

The problem of mirror moving in the quantum vacuum is in many respects
similar to the polaron problem in condensed matter: a single particle
interacting with the quantum field of phonons or other bosonic field.
In the polaron problem the Hamiltonian for quantum  particle of
bare mass $M$ interacting with the scalar field (annihilation operator
$c_{\bf q}$) has the following form, see e.g. review \cite{Devreese2009}:
\begin{equation}
H= -\frac{\nabla^2}{2M} +
 \sum_{\bf q}V_{\bf q}(c_{\bf q} + c^+_{-{\bf q}})e^{i{\bf q}\cdot{\bf r}}
+ \sum_{\bf q} \omega({\bf q}) c^+_{{\bf q}}c_{\bf q}
\,.
\label{HamiltonianPolaron}
\end{equation}
As distinct from the conventional polaron problem, in the mirror
problem in Ref. \cite{WangUnruh2013}
the interaction $V({\bf q})$ with the scalar field is mediated by the
oscillator.
The quantization of
the classical action Eq. (\ref{ClassicalAction}) gives rise to the following
polaronic-like  Hamiltonian :
\begin{eqnarray}
H= - \frac{\nabla^2}{2M}
+ \Omega \left(b^+b +\frac{1}{2} \right)+
\sum_q \omega(q) c^+_{q}c_q +
\label{HamiltonianMirror}
\\
+ \epsilon\sum_q \frac{\sqrt{\Omega}}{2\sqrt{\omega(q)}}(b-b^+)(c_q +
c^+_{-q})ie^{iqx}+
\label{HamiltonianMirrorCorrection1}\\
+\sum_{q,q'} \frac{\epsilon^2}{4\sqrt{\omega(q)\omega(q')}}(c_q
+c^+_{-q})(c^+_{q'}
+c_{-q'})e^{i(q-q')x} \,.
\label{HamiltonianMirrorCorrection2}
\end{eqnarray}
Here $b^+= \sqrt{\Omega/2}(q -\frac{1}{\Omega}\partial_q)$ is the creation
operator for the oscillator quanta; the second order in $\epsilon$ term in
Eq.(\ref{HamiltonianMirrorCorrection2}) is $\frac{1}{2}\epsilon^2\phi^2$. It
appears due to the transformation $H=L-\dot q \partial L /\partial \dot q$
from the Lagrangian description in terms of velocity
 $\dot q$ to the Hamiltonian description in terms of momentum $p=\dot
q-\epsilon \phi=-i\partial_q$.

\subsection{Effective mass of the mirror}
\label{SectMnr}

The perturbation correction to the energy of polaron in
Eq.(\ref{HamiltonianPolaron}) at zero momentum is \cite{Devreese2009}:
\begin{equation}
\Delta {\cal E}= - \sum_{\bf q} \frac{|V_{\bf q}|^2}{\omega({\bf q})  +
\frac{q^2}{2M}}
\,.
\label{MassCorrectionPolaron}
\end{equation}
The negative correction comes from the conventional second order
perturbation theory,
where the excited states are formed by
particles with momentum ${\bf q}$ and phonons with momentum $-{\bf q}$.

In case of a mirror one has two $\epsilon^2$ contributions: from the
second order perturbation theory in
Eq.(\ref{HamiltonianMirrorCorrection1}) with $V_q = \epsilon
\frac{\sqrt{\Omega}}{2\sqrt{\omega(q)}}(b-b^+)$.
\begin{eqnarray}
\Delta  {\cal E}_1 &=& - \sum_{q} \frac{\langle V_{q}^+ V_q\rangle}{\Omega +
\omega(q)  +
\frac{q^2}{2M}}\nonumber\\
&=&- \frac{\epsilon^2\Omega}{4\pi} \int_0^\infty   \frac{dq}{\omega(q)}\,
\frac{1}{\Omega
+\omega({q})  + \frac{q^2}{2M}}
\,,
\label{FirstMassCorrection}
\end{eqnarray}
 and from the first order perturbation theory for perturbation in
Eq.(\ref{HamiltonianMirrorCorrection2}):
\begin{equation}
\Delta  {\cal E} _2 = \frac{1}{2} \epsilon^2  \left<\phi^2\right>=
\frac{\epsilon^2}{4\pi} \int_0^{\infty}  \frac{dq}{\omega({q})} \,.
\label{E2}
\end{equation}
The negative term Eq. (\ref{FirstMassCorrection}) corresponds to the
conventional correction to the polaronic energy in
Eq.(\ref{MassCorrectionPolaron}), but now the excited state includes also
the first excited level of the oscillator. The positive term Eq. (\ref{E2})
follows from the dependence of the interaction on the time derivative $\dot
q$.

Altogether one has
\begin{eqnarray}
\Delta {\cal E}  \approx   \epsilon^2\,  \int_0^{\infty} \frac{dq}{4\pi
\omega(q)} \Bigl(1 - \frac{\Omega}{\Omega +\omega(q)  + \frac{q^2}{2M}}
\Bigr)
\label{E}
\end{eqnarray}
Thus
the total mass of the mirror in
the second order approximation in $\epsilon$ is:
\begin{equation}
M_{\rm eff}=M +\frac{\Omega}{2} +  \epsilon^2\,  \int_0^{\infty}
\frac{dq}{4\pi \omega(q)}
\Bigl(1 - \frac{\Omega}{\Omega +\omega(q)  + \frac{q^2}{2M}}
\Bigr)
\label{TotalMass}
\end{equation}
In case of a  large bare mass of mirror $M\gg \Omega\gg \epsilon^2$ and for
$\omega(q)=q$, the effective mass of the mirror is:
\begin{eqnarray}
M_{\rm eff}=M +\frac{\Omega}{2}+  \frac{\epsilon^2} {4\pi} \ln
\frac{E_{\rm UV}}{\Omega}\,,
\label{LargeMassApprox}
\end{eqnarray}
where $E_{\rm UV}$ is the UV cutoff. We consider the relativistic spectrum
$\omega(q)=q$ of the scalar field, but the non-relativistic limit for the
mirror, i.e. the condition $q^2/2M\ll q\ll M$. This means that
Eq.(\ref{LargeMassApprox})
is valid if
\begin{equation}
M\gg E_{\rm UV}\gg \Omega\gg \epsilon^2\,.
\label{NonrelativisticCondition}
\end{equation}
For the smaller mass $M$ one should consider the relativisictic description
of the mirror motion.

Eq.(\ref{LargeMassApprox}) coincides with Eq. (111) in Ref.
\cite{WangUnruh2013} in the limit
$\Omega\gg \epsilon^2$ with logarithmic accuracy (in Ref.
\cite{WangUnruh2013} the recoil energy $k^2/2M$ has been neglected, which
assumes the condition $M\gg \Omega$). The non-logarithmic difference  of the
order of $\epsilon^2$ between the results can be attributed to the fact that
we considered
the  quantum limit for the oscillator. In  Ref. \cite{WangUnruh2013} the
classical equation for the oscillator variable has been used, which is more
appropriate in the limit
$\Omega\ll \epsilon^2$.  In this limit Eq. (111) in Ref.
\cite{WangUnruh2013} gives
\begin{eqnarray}
M_{\rm eff}=M  +  \frac{\epsilon^2} {4\pi}
\ln \frac{E_{\rm UV}}{\epsilon^2}\,.
\label{SmallOmegaApprox}
\end{eqnarray}

\subsection{Regularization due to the finite size of the mirror}
\label{UVsection}

Both results for the effective mass of the mirror in Eqs.
(\ref{LargeMassApprox}) and
(\ref{SmallOmegaApprox}) contain the UV  diverging logarithm.
  The UV divergence comes form the $\delta$-functional interaction of the
mirror with the scalar field, which corresponds to the approximation of the
infinitely thin mirror. The strong localization of the oscillator in space
causes the UV divergence at  $k\rightarrow \infty$.

To cure this divergence one should consider the
more realistic smeared interaction  $\epsilon U_{\rm nr}(x)$ between the mirror and
the scalar field instead of the sharp interaction $\epsilon \delta(x)$ in Ref. \cite{WangUnruh2013}.
The  form-factor $U_{\rm nr}(x)$ is localized at the finite distances
$\sim R$, which corresponds to the finite
width of the mirror. This will be done rigorously in Section \ref{SectR},
where the relativistic case is considered. The non - relativistic case of
the present section may be obtained as a corresponding limit of the
expressions of Sect. \ref{SectR}.  The form-factor $U_{\rm nr}(x)$ results
in the renormalization of the scalar field propagator (see Eq.
(\ref{renorm})). As a result in the non - relativistic case we arrive at
\begin{eqnarray}
\Delta M  = \frac{\epsilon^2}{4\pi} \int_0^\infty  \, dk~ \frac{|U_{\rm nr}(k)|^2}{\omega(k)}\,
\Bigl(1 - \frac{\Omega}{\Omega +\omega(k)  + \frac{k^2}{2M}}
\Bigr)\,.
\label{MassCorrectionCutoff}
\end{eqnarray}
Here $U_{\rm nr}(k)$ is the Fourier transform of $U_{\rm nr}(x)$.
So, the UV cut-off in Eqs. (\ref{LargeMassApprox})
and
(\ref{SmallOmegaApprox}) is provided by the characteristic length scale   of
the
potential $U_{\rm nr}(x)$,
i.e.  $E_{\rm UV}\sim 1/R$.
The condition (\ref{NonrelativisticCondition}) becomes:
\begin{equation}
M \gg \frac{1}{R}\gg  \Omega \gg \epsilon^2\,.
\label{NonrelativisticCondition2}
\end{equation}
This means that the thickness of the mirror must be larger than its Compton
wavelength.

\section{Relativistic moving oscillator}
\label{SectR}

\subsection{Field - theoretical description}

 Let us start from the action given in \cite{WangUnruh2013} for the moving
oscillator interacting with the scalar field written in a different form:
\begin{eqnarray}
S &=&  \int d\tau\Bigl(-M + \frac{1}{2} \dot{q}^2(\tau) - \frac{\Omega^2}{2}
q^2(\tau) -
\epsilon \, \dot{q}(\tau)  \phi(y[\tau]) \Bigr)
\nonumber\\&&+
\frac{1}{2}\int\!d^2 x\,
\partial_i{\phi}\partial^i \phi
\label{S0}
\end{eqnarray}
Here $y[\tau]$ is the trajectory of the particle, $\tau$ is the proper time.
This action describes the relativistic particle with mass $M$ and the
oscillator inside it moving along the trajectory $y^i[\tau]$.

The field - theoretical description for this moving object gives the
partition function
\begin{eqnarray}
Z_1& =& \int D y[\tau] Dq[\tau] \, {\rm exp}\Bigl(i  \int d\tau \Bigl(-M +
\frac{1}{2} \dot{q}^2(\tau) - \frac{\Omega^2}{2} q^2(\tau) \Bigr) \nonumber\\&&- i
\epsilon \int \dot{q}(\tau) \phi(y(\tau)) d\tau \Bigr),
\end{eqnarray}
where the integral is over the world trajectory of the particle $y$ and over
the oscillator coordinates $q$. Let us first work out the integral over $q$.
We denote $
f(\tau) = \epsilon \phi(y(\tau))$.
The
standard methods allow to rewrite
\begin{eqnarray}
Z^{(0)}_1[y(\tau)] &=& \int Dq[\tau] \, {\rm exp}\Bigl(i  \int d\tau
\Bigl(-M + \frac{1}{2} \dot{q}^2(\tau) \nonumber\\&&- \frac{\Omega^2}{2} q^2(\tau) -
\dot{q}(\tau) f(\tau) \Bigr)\Bigr)\nonumber\\
&=&{\rm Tr} \, P\, {\rm exp} (-i \int d \tau \hat{H}(\tau)),
\end{eqnarray}
where
\begin{equation}
\hat{H}(\tau) =  M + \frac{\Omega^2}{2} q^2   +  \frac{1}{2}( - i \frac{d}{d
q} )^2 +    ( - i \frac{d}{d q} ) f(\tau) + \frac{1}{2} f^2(\tau)
\end{equation}
Next, we introduce the annihilation operator
\begin{equation}
a = \sqrt{\frac{\Omega}{2}}\Bigl(q + \frac{1}{\Omega}\frac{d}{d q} \Bigr)
\end{equation}
This gives
\begin{equation}
\hat{H}(\tau) =  M + \Omega (a^+ a + 1/2) - i \sqrt{\frac{\Omega}{2}}
(a-a^+)f(\tau) + \frac{1}{2}f^2(\tau)
\end{equation}

Also we introduce the operators $\hat{N} = a^+a$ and $\hat{P} = i (a -
a^+)$, and the dimensionless constant $\hat{\epsilon} =
\epsilon/\sqrt{2\Omega}$.  Now partition function $Z_1$ describes the
particle with the internal discrete degree of freedom $n = 0,1,2,...$
\begin{eqnarray}
Z_1 &=&\int Dy(\tau) {\rm Tr} \, P\, {\rm exp} \, \Bigl[-i \int d \tau
\Bigl( M\nonumber\\&& + \Omega (\hat{N} + 1/2 - \hat{\epsilon}  \hat{P}  \phi(y(\tau)) +
\hat{\epsilon}^2 \phi^2(y(\tau))  \Bigr) \Bigr]
\end{eqnarray}
We define mass of the moving oscillator at the ground level $|0\rangle$:
$\tilde{M} = M + \Omega/2$. It takes into account the contribution of the
oscillator with $n=0$. The corresponding relativistic field theory is
described by the multi - component spinor $\Psi_a$, $ a = 0,1,2,...$.
If the given particle is fermionic, the corresponding partition funciton
receives the form:
\begin{eqnarray}
Z &=& \int\!\!D \bar{\Psi}(x,t) \, D \Psi(x,t) \,D \phi(x,t)\, \nonumber\\&&{\rm
exp}\Bigl(\frac{i}{2}\!\!\int\!\!dx\,dt\,
\Big[
{(\partial_t\phi)^2}
- {(\partial_x\phi)^2}\Big] \nonumber\\ && + i\!\!\int\!\!d x\,dt\,
\bar{\Psi} [  i\partial_k  \gamma^k - \tilde{M} \nonumber\\&&- \Omega (\hat{N}  -
\hat{\epsilon}  \hat{P}  \phi(x) + \hat{\epsilon}^2 \phi^2(x) )  ] \Psi
\Bigr),
\label{ftfrohlich3_}
\end{eqnarray}
Here the sum is over $k = 0,1$, and $\gamma^0 = \sigma^1, \gamma^1 = i
\sigma^2$. The grassmann variable $\Psi$ is multi - component. In practise
we may consider the $K$ - component spinor with $K \gg 1$. Coupling constant
in this theory is dimensionless. Therefore, the divergences are at most
logarithmic.

\subsection{An alternative formulation}

To check that the one - particle action of Eq. (\ref{S0}) indeed corresponds to the second - quantized system with partition function of Eq. (\ref{ftfrohlich3_}) let us come to the latter formulation by an alternative way. Namely, let us shift the derivative over $\tau$ in  the term $\dot{q}(\tau)  \phi(y[\tau])$ of Eq. (\ref{S0}) to $\phi$:  $\dot{q}(\tau)  \phi(y[\tau])\rightarrow -q(\tau) \frac{d}{d\tau} \phi(y[\tau])$. The second - quantized version of the theory for this action looks different from that of given by Eq. (\ref{ftfrohlich3_}).
It
gives the following partition function
\begin{eqnarray}
Z &=& \int\!\!D \bar{\Psi}(x,t) \, D \Psi(x,t) \,D \phi(x,t)\, \label{ftfrohlich3_1}\\&&{\rm
exp}\Bigl(\frac{i}{2}\!\!\int\!\!dx\,dt\,
\Big[
{(\partial_t\phi)^2}
- {(\partial_x\phi)^2}\Big] \nonumber\\ && + i\!\!\int\!\!d x\,dt\,
\bar{\Psi} [ ( i\partial_k  - \hat{\epsilon} \partial_k \phi(x)
\hat{Q} )\gamma^k - \tilde{M} - \Omega \hat{N}  ] \Psi
\Bigr),
\nonumber
\label{ftfrohlich3_1}
\end{eqnarray}

This looks like the system of spinor field in the presence of a very specific gauge field  $\hat{\epsilon} \partial_k \phi(x)
\hat{Q}$. The action does not contain the interaction term with the second power of the field $\phi$. However, we shall prove below that this system is indeed equivalent to that of Eq. (\ref{ftfrohlich3_}).
Let us apply the following gauge transformation
\begin{equation}
\Psi(x) \rightarrow e^{i \hat{\epsilon} \hat{Q} \phi(x)}\Psi(x)
\end{equation}

This results in
\begin{eqnarray}
Z &=& \int\!\!D \bar{\Psi}(x,t) \, D \Psi(x,t) \,D \phi(x,t)\, \nonumber\\&&{\rm
exp}\Bigl(\frac{i}{2}\!\!\int\!\!dx\,dt\,
\Big[
{(\partial_t\phi)^2}
- {(\partial_x\phi)^2}\Big] \nonumber\\ && + i\!\!\int\!\!d x\,dt\,
\bar{\Psi} [  i\partial_k \gamma^k - \tilde{M} - \Omega \hat{R}  ] \Psi
\Bigr),
\label{ftfrohlich3_2}
\end{eqnarray}
where
\begin{equation}
\hat{R} = e^{-i \hat{\epsilon} \hat{Q} \phi(x)}\hat{N}e^{i \hat{\epsilon}
\hat{Q} \phi(x)}
\end{equation}
Using properties of the coherent state displacement operator $D(\alpha) =
e^{\alpha a^+ - \alpha^* a}$ we can prove, that
\begin{eqnarray}
\hat{R} &=& e^{-i \hat{\epsilon} \hat{Q} \phi(x)}\hat{N}e^{i \hat{\epsilon}
\hat{Q} \phi(x)}
=  \hat{N} - \hat{\epsilon} \hat{P} \phi + \hat{\epsilon}^2 \phi^2
\end{eqnarray}
This gives us the partition function in the form of Eq.
(\ref{ftfrohlich3_}).

\subsection{Mass of the moving oscillator}

Here and below in this subsection we use Euclidean formulation of the model, i.e. assume that the Wick rotation is performed. The  one - loop perturbation theory relates the
correction to the mass of the moving oscillator $\Delta {\cal E}$ with the
self - energy function $\Sigma(p)$.
Let us calculate this function using the formulation of Eq.
(\ref{ftfrohlich3_}).  We have $\Sigma(p) = \Sigma^{(2)} + \Sigma^{(1)}$
with
\begin{eqnarray}
\label{polaron1}
\Sigma^{(1)}(p)   &\approx & i \,  \hat{\epsilon}^2\, \Omega  \int \frac{d^2
k }{(2\pi)^2}\frac{1}{k^2}
\end{eqnarray}
This term corresponds to the one - loop diagram caused by the term
$\hat{\epsilon}^2 \Omega \phi(x)^2$.
The diagram originated from the term $\hat{\epsilon}^2 \hat{P}^2
\phi(x)\phi(y)$ gives:
\begin{eqnarray}
\label{polaron2}
\Sigma^{(2)}(p)   &\approx &  -\, \hat{\epsilon}^2\,  \Omega^2 \int \frac{d^2
k }{(2\pi)^2}\nonumber\\&&\langle0|
\hat{P}\frac{1}{(p+k) \sigma - i\tilde{M} - i\Omega
\hat{N}}\hat{P}\frac{1}{k^2}|0\rangle
\end{eqnarray}

Here $|0\rangle$ is the ground state of the oscillator.
We have
\begin{equation}
\label{polaron3}
\Sigma^{(2)}(p)
\approx  -\, \hat{\epsilon}^2\,  \Omega^2 \int \frac{d^2 k }{(2\pi)^2k^2}
\frac{1}{(p+k) \sigma - i\tilde{M} - i\Omega }
\end{equation}

The non - relativistic limit was described in the previous section. It may be obtained here if we suppose that the UV cut - off $\Lambda \ll M$. Then integration over $k^0$ gives us the expressions of Sect. \ref{SectMnr}.
Here we consider the opposite limit, when the UV cutoff $\Lambda \gg M$.
We get
\begin{eqnarray}
&&\Sigma^{(1)}(p)+\Sigma^{(2)}(p)
\approx  \, i \hat{\epsilon}^2\,  \Omega \int \frac{d^2 k }{(2\pi)^2k^2}
\nonumber\\&&\frac{((p+k) \sigma - i\tilde{M})((p+k) \sigma + i\tilde{M} +
i\Omega )}{(p+k)^2 + (\tilde{M} +\Omega)^2 }
\end{eqnarray}

Since near the pole $p\sigma - i\tilde{M} \sim \epsilon^2$, we may set $p^2
= - \tilde{M}^2$ and neglect the terms proportional to
$p\sigma - i\tilde{M}$ as these terms result in the renormalization of the
propagator and do not contribute to the renormalized mass. The remaining
terms are given by
\begin{eqnarray}
\Sigma^{\prime}(p)
&\approx & \,\, i \hat{\epsilon}^2\,  \Omega \int \frac{d^2 k }{(2\pi)^2k^2}
\frac{k^2 + (2+\frac{\Omega}{\tilde{M}}) (kp)}{(p+k)^2 + (\tilde{M} +\Omega)^2 }
\end{eqnarray}
and are related to the correction to the renormalized mass of the mirror
$M_R$ as $\Sigma^{\prime}(i\tilde{M},0) = i (M_R-\tilde{M})$.
The direct calculation of the integral gives
\begin{eqnarray}
M_R &=&
 {M} + \Omega/2 + \frac{ {\epsilon}^2}{4\pi} {\rm
log}\Bigl[\frac{\Lambda}{\Omega}\,\zeta\Bigl(\frac{\Omega}{2{M}+\Omega}\Bigr
)\Bigr] \label{UVDM}
\end{eqnarray}
Here
\begin{equation}
\zeta(x) = \frac{1}{1+\frac{1}{2x}}\Bigl(1+\frac{1}{4\Bigl(x^2 +
x\Bigr) }\Bigr)^{1+x}
\end{equation}
This function varies between $1$ and $1/2$. Therefore, at $\Lambda \gg
\Omega$ we estimate
\begin{eqnarray}
M_R & \approx &  {M} + \Omega/2 + \frac{ {\epsilon}^2}{4\pi} {\rm
log}\frac{\Lambda}{\Omega} \label{UVDM_}
\end{eqnarray}
for any relation between $M$ and $\Omega$. This coincides with the non -
relativistic result of Eq. (\ref{LargeMassApprox}).
However, in this case we imply, that
\begin{equation}
\frac{1}{R}\gg M, \Omega \gg \epsilon^2\,
\label{NonrelativisticCondition3}
\end{equation}
This means that the thickness of the mirror $R$ must be much smaller than
its Compton
wavelength.

\subsection{Regularization due to the finite size of the mirror}
\label{UVRsection}

Let us introduce the form - factor $U(t,x)$ to regularize the UV divergence that comes form the original $\delta$-functional interaction between mirror and scalar field.
The corresponding action is given by
\begin{eqnarray}
S&=& \frac{1}{2}\!\!\int\!\!d^2x\,
\Big[
{(\partial_t\phi)^2}
- {(\partial_x\phi)^2}\Big] \nonumber\\ && + \!\!\int\!\!d^2 x\,
\bar{\Psi} [  i\partial_k  \gamma^k - \tilde{M} - \Omega \hat{N} ] \Psi
\nonumber\\&& + \Omega\hat{\epsilon} \!\!\int\!\!d^2 x\,
\bar{\Psi}(x)   \hat{P} \Psi(x) \int \, d^2z\, U(x-z) \phi(z)  \nonumber\\&& - \Omega \hat{\epsilon}^2 \!\!\int\!\!d^2 x\,
\bar{\Psi}(x)   \Psi(x)  \Big(\int \, d^2z\, U(x-z)\phi(z)\Big)^2
\label{ftfrohlichCorrectedR}
\end{eqnarray}
The model considered in Ref. \cite{WangUnruh2013} is restored in the
infinitely thin limit   $U(x,t)=\delta(x)\delta(t)$.
One can see, that the effect of the form - factor $U$ may be taken into account via the renormalization of the scalar field propagator
\begin{equation}
\frac{1}{k^2} \rightarrow U(k) \frac{1}{k^2} U(-k)\label{renorm}
\end{equation}
Here $U(k) = U(k_0,k_1)$ is the Fourier transform of $U(t,x)$. The limit of the infinitely thin mirror appears when $U(k) = 1$.
In the non - relativistic case we should consider $U(t,x) = U_{\rm nr}(x)\delta(t)$ and assume, that $U_{\rm nr}$ is real - valued. In the general case of relativistic - invariant theory the situation is more involved. The form - factor $U$ (that may be complex - valued) should depend on the invariant interval $s^2 = t^2-x^2$ and decreases fast at $|s^2|\rightarrow \infty$. That means that the mirror moving along the particular world trajectory interacts with the scalar field along the whole light cone. Therefore, in relativistic case we do not interpret the form - factor in terms of the arbitrary world trajectories of the particles.

In non - local relativistic quantum field theory the finite size of the source may be taken into account using the form - factors in momentum space that depend on the invariant $p^2$ (see \cite{Efimov} and references therein). At least, for the static sources of the finite size such form - factors indeed model the field caused by the charge distributed over the finite region of space. Below we consider the particular form of $U$ that is caused by the interaction with an intermediate auxiliary field $\theta$ of mass $\Lambda$. Point - like mirror emits/absorbs one or two quanta of the field $\theta$.  The vertexes for the emission/absorbsion of $\theta$ are the same as those for the emission/absorbsion of $\phi$ in Eq. (\ref{ftfrohlich3_}). The quanta of the field $\theta$ may be transformed to the quanta of the field $\phi$. We know that the exchange by massive particles in field theory describes the interaction that occurs at finite distances. This finite distance may be evaluated as $R = 1/\Lambda$. That's why we consider the exchange by massive particle as the device for modeling the finite size of the mirror. This leads us to the interpretation of $U$ as the propagator of $\theta$ (up to the normalization constant) while Eq. (\ref{ftfrohlichCorrectedR}) is the effective action of the theory obtained after the field $\theta$ is integrated out.
This results in the following form of the function $U(k)$ in momentum space
\begin{equation}
U(k) = -\frac{\Lambda^2}{k^2 - \Lambda^2 + i 0},
\end{equation}
The corresponding function $U(t,x)$ in $1+1$ D coordinate space is expressed through the complex - valued special functions. One can check, that the absolute value of $U(t,x)$ falls sharply at $|t^2-x^2|\Lambda^2 \rightarrow \infty$.    For the renormalized mass of the mirror we have
\begin{eqnarray}
M_R &=&
 {M} + \Omega/2 \nonumber\\&&+  \hat{\epsilon}^2 \Omega  \int \frac{d^2 k }{(2\pi)^2k^2}U(k)U(-k)
\frac{k^2 + (2+\frac{\Omega}{\tilde{M}}) (kp)}{(p+k)^2 + (\tilde{M} +\Omega)^2 }\label{UVDMR}
\end{eqnarray}
Here the integral is over the Euclidean $2$ - momentum $k$, while $U(k) = \frac{1}{1+k^2 R^2}$ is obtained from $U(k_0,k_1)$ by analytical continuation. Then one can easily derive (at $\Omega R \ll 1$)
\begin{eqnarray}
M_R & \approx &  {M} + \Omega/2 + \frac{ {\epsilon}^2}{4\pi} {\rm
log}\frac{1}{\Omega R} \label{UVDM_R}
\end{eqnarray}

\section{Conclusion}

We suggest the description of the model for the $1+1$ D mirror discussed in
Ref. \cite{WangUnruh2013} that reveals its analogy to polaron. We consider
the field theory that contains the multi - component spinor field
interacting with the scalar field. This is the second quantized theory that
describes moving oscillator interacting with the scalar field. Actually,
this model may easily be formulated in space - time of any dimension. Being
defined in $3+1$ D it may have certain applications in the high - energy
physics if it is necessary to describe the fermionic particle with infinite
(or, large) number of internal energy levels. (For example, in \cite{bodo}
it is suggested, that such an internal degree of freedom marks the flavor of
the Standard Model fermion.) However, here we restrict ourselves by the
consideration of the $1+1$ D case that exactly matches the model of
\cite{WangUnruh2013}. In this case the  moving oscillator may be considered as a model of moving mirror.

We consider the two versions of the second - quantized theory of the mirror. The two formulations are exactly equivalent as follows from our analysis. The formulation of Eq. (\ref{ftfrohlich3_}) reveals the analogy with the usual polaron problem. At the same time the formulation of Eq. (\ref{ftfrohlich3_1}) contains the interaction of the mirror with the special gauge field composed of the scalar excitations. We suppose, that this formulation may be useful for the further applications of the developed formalism to various problems both in the high energy physics and in the condensed matter physics.

We demonstrate, that in the discussed model there appears the extra term
(Eqs. (\ref{E2}), (\ref{polaron1}))  in
the energy of the polaron, which is logarithmically divergent. This extra
term reverses
the sign of the energy correction as compared to the negative mass
correction (Eqs.
(\ref{MassCorrectionPolaron}), (\ref{polaron3})) in the conventional
polaron problem.

The obtained infinite value of the mirror
mass is the artifact of the model, which uses the artificial
$\delta$-function potential. The logarithmically divergent integral is
regularized by any realistic potential which provides the natural UV
cut-off. In this mirror-polaron problem,  the UV
divergence is the
physical effect, while different magnitudes $E_{\rm UV}$ of the UV
cut-off reflect different physical mechanisms.

On the contrary, in our opinion,  in the cosmological constant
problem the $k^4$ UV
divergence of the zero point energy of quantum field is unphysical. It has been suggested in \cite{VolovikKlinkhamer2008,Volovik2013} that for the consideration of this problem the variable overall volume of the Universe is to be taken into account. As a result, in the Unverse near to the equilibrium the vacuum energy is nearly zero. This points out that there should be the contributions to the vacuum energy coming from the unknown physics at the energies above the cutoff of the Standard Model, that are to cancel exactly the mentioned divergencies. If so, the counting of zero
point energies of the low energy effective theory for the calculation of the vacuum energy density cannot be applied.
This follows the analogy with the condensed matter systems,  where the exact microscopic "trans-Planckian" theory is known, while the energy density of the
macroscopic ground state  is not determined by the quantum fluctuations or by the microscopic physics. Instead it is determined by the environment. For the vacuum in a full thermodynamic equilibrium and in the absence of environment, the vacuum energy density is exactly zero
\cite{VolovikKlinkhamer2008,Volovik2013}. This is the consequence of the
integrated form of the Gibbs-Duhem equation, which is generic and is applicable both to the relativistic vacuum and to the non-relativistic condensed matter systems. This means in the equilibrium vacuum of condensed matter system the zero point contributions to the vacuum energy are completely canceled by the microscopic contributions.

The work of M.A.Z. is  supported by the Natural Sciences and Engineering
Research Council of
Canada. GEV
acknowledges a financial support of the Academy of Finland and its COE
program.

\end{document}